# Capital Structure Theories and its' Practice:

## A study with reference to select NSE listed public sectors banks, India

**Kurada T S S Satyanarayana[1]\*, Addada Narasimha Rao[2,]\***

[1]Junior Research Fellow, Department of commerce and Management studies, Andhra University,

[2]Professor, Department of commerce and management studies, Andhra University.

Emails: [1] satyanarayanaktss.rs@andhrauniversity.edu.in

[2] trishulkurada963@gmail.com

## Abstract

Among the various factors affecting the firm's positioning and performance in modern day markets, capital structure of the firm has its own way of expressing itself as a crucial one. With the rapid changes in technology, firms are being pushed onto a paradigm that is burdening the capital management process. Hence the study of capital structure changes gives the investors an insight into firm's behavior and intrinsic goals. These changes will vary for firms in different sectors. This work considers the banking sector, which has a unique capital structure for the given regulations of its operations in India. The capital structure behavioural changes in a few public sector banks are studied in this paper. A theoretical framework has been developed from the popular capital structure theories and hypotheses are derived from them accordingly. The main idea is to validate different theories with real time performance of the select banks from 2011 to 2022. Using statistical techniques like regression and correlation, tested hypotheses have resulted in establishing the relation between debt component and financial performance variables of the select banks which are helping in understanding the theories in practice.

## 1 Introduction

For a firm, setting a newer or higher objective will always lead to the necessity of making some potential changes in its existing level of the capital. Additionally, in a corporate environment where the idea of wealth maximisation is what motivates an organization's financial management strategies to maintain, gain, and enhance its worth, changes in capital mix will unavoidably have an effect on how the market perceives the company [1]. Financing mix is a composition of long-term resources like equity, preference shares, debentures, bonds and other long-term loans [2]. Different sources of finance have various characteristics of risk-return. Determination of effective capital mix is crucial for mitigating firms' financial risk and reducing the cost of capital. Since, Capital structure decisions are important for firm's survival and growth, there has been an evolution of extensive studies in this area. These are more concerned on matters like cost-benefit analysis, sudden threats demanding firm's capital structure changes, and reluctance offered by the system while deciding on switching the capital mix [3, 4]. While other research have focused on determining the variables that affect capital structure decisions or, conversely, how capital structure policy affects the performance of the business, the goal of the current work is to determine the firm's behaviour based on the accepted theories. Prime factors, affecting the firm behaviour are vital for generating enhanced customer satisfaction. One of these elements that puts stress on capital management is technology. Although it has an influence on every industry, this study on the Indian banking industry seeks to uncover business behaviour grounded in capital structure theories. The choice was made because the banking industry is being negatively impacted by rapid technological change.

A country's banking industry plays a significant part in its social and economic growth. Making decisions on these banks' financial features includes choosing a capital structure that sets them apart from other companies and non-banking institutions. The theoretical and empirical relationship between financial performance and capital structure remains an undetermined problem [5]. The importance of capital structure is prioritised in order to raise the value of the bank, the market value of shares and securities, and to safeguard banks from both over and undercapitalization.

The theoretical underpinning of the present study, variables of interest, the development of hypotheses, and the justification for choosing a group of specified public sector banks are discussed in the methodology sections that follow a brief review of the capital structure theories employed to develop it. At last Section 4 covers the empirical model and data used to conduct the analysis; results are discussed in section 5.

## 2 Review of capital structure theories

Capital structure is a methodical way to finance corporate operations using a mixture of both equity and debt. The relationship between equity financing and debt financing is examined by a list of capital structure theories that have been compiled in the literature. These theories include well-known concepts like the Net Income Theory, Net Operating Income Theory, Traditional Theory, Modigliani and Miller Theory, Pecking Order Theory, and Agency Cost Theory. This work is intended to generalize the relationship established between the capital structure variables from the above theories and thus create a model framework that could help in tracing the firm's capital structure policy.

The net income theory of capital structure suggested by David Durand (1950), is a financial approach that contends the capital structure that maximises a company's net income is its ideal capital structure. According to this idea, a company should base the combination of equity and debt that it uses to fund its operations on the predicted net income it would produce by utilising different amounts of equity and debt. In accordance with this principle, a business should decide to employ more debt financing if doing so increases net income. On the other hand, if additional debt leads to a decrease in net income, the firm should reduce its use of debt and increase its use of equity financing [6].

The net operating income (NOI) theory of capital structure was proposed by David Durand (1952), is a financial theory that argues that a firm's optimal capital structure is the one that maximizes its net operating income. It states that the choice of equity and debt for a company depends on the anticipated net operational income it produces by utilising different kinds of both equity and debt. A company should employ more debt financing up till its net operating revenue increases. The company must lower the debt component if any extra debt is lowering its NOI. The NOI theory of capital structure is a more focused approach than the net income theory, as it considers only the operating income generated from a firm's core business operations and ignores non-operating items such as gains from investments or changes in taxes [7].

The traditional theory of capital structure, also known as the trade-off theory, is a financial theory that argues that a firm's optimal capital structure is a trade-off between the benefits of using debt financing, such as lower cost of capital and tax benefits, and the costs of using debt financing, such as financial risk and bankruptcy costs. According to this theory, a firm should strive to achieve an optimal balance between debt and equity financing, considering the trade-off between the benefits and costs of each financing option. The traditional theory suggests that as a firm increases its use of debt financing, it becomes more financially leveraged, which increases its financial risk and the risk of bankruptcy. The conventional approach also contends that as a corporation uses debt financing more frequently, its cost of capital lowers and its tax advantages rise, which can enhance net income. Financial managers frequently utilise the classical theory of capital structure when deciding which investments to make since it gives an outline for understanding the link within debt and equity financing [8].

MM Theory (1958), named after Franco Modigliani and Merton Miller, is a financial theory that argues that a firm's capital structure does not affect its value in a perfect market. This theory contends that a company's market value is derived by its assets and earnings and is unaffected by the proportion of equity and debt financing used to fund those assets. According to the MM theory, regardless of a firm's capital structure, the cost of capital is the same in a perfect market without taxes, transaction fees, or information asymmetry. As a result, the MM theory argues that firms should choose the capital structure that is optimal for them, considering their unique circumstances and constraints, rather than trying to optimize their capital structure to maximize value. The MM theory is generally recognised in financial theory and offers a helpful framework for comprehending the connection involving capital structure and business value [9].

The pecking order theory (1984) of capital structure, proposed by Mayer's and Majluf is a financial theory that argues that firms have a preferred order of financing sources, with internal financing (retained earnings) being the first choice, followed by debt financing, and finally, equity financing. This theory holds that corporations seek to finance their investment decisions using internal funds as these funds are the least costly and have the least knowledge asymmetry. If internal funds are not available, firms will then turn to debt financing, as debt is considered less risky than equity financing and has a lower cost of capital. If debt financing is not feasible or desirable, firms will then turn to equity financing as a last resort, as equity financing has the highest cost of capital and is considered the riskiest form of financing. The pecking order hypothesis, which is generally recognised in financial theory, offers a helpful framework for comprehending the financing behaviour of businesses [10].

The agency costs theory of capital structure is a financial theory that argues that a firm's capital structure affects its value through the agency costs associated with different forms of financing. This theory contends that the use of financing through debt raises agency costs by introducing ethical hazards and the possibility of managers acting opportunistically. The theory argues that debt financing creates a "debt overhang" that can limit a firm's flexibility and constrain its ability to invest in positive-NPV projects. On the other side, equity financing generates agency costs since it reduces current shareholders' ownership and control. According to the agency costs hypothesis, businesses should aim to strike an ideal equilibrium between equity and debt funding in order to reduce the agency costs related to each kind of financing. [11].

## 3   Theoretical Framework

As the above theories have laid a foundation to understand the relationship between the capital structure and firms positioning, the objective of the present section is to derive propositions that suits well in reading the firm's capital structure behaviour in real time business environment. Firms positioning, is referred to two main aspects, market value or perceived value of the firm by investors and the other is its overall performance. All the theories proven to be working are developed under certain assumptions, which are more likely to be regarded as ideal and impractical. Some of the impractical assumptions among them relating to capital structure theories may be like assuming a perfect market condition, availability of fully symmetrical information, excluding the tax impact etc.

From these theories, there is an opportunity found to generalize the trade-off between different capital structure indicators. For this sake, stringent assumptions of the theories were kept aside, and correlated aspects are aimed to be directly tested in practical environment.

This work considers the capital mix problem of a listed firm in stock markets. The moment a firm have entered the equity markets to raise the capital, it is generally deemed to have utilised the other sources of financing like debentures, loans etc. It is the point where the investors start the assessment of market value, there by becoming a vital factor in influencing the firm's behaviour [12]. The scanning of the firm's intrinsic behaviour, through leverage, profitability ratios, EBIT-EPS analysis are generally known to be in the basket of investors. Adding the capital structure analysis would give knowledge in making more accurate decisions.

To perform capital structure behavioural analysis, a three-fold procedure is being proposed, with step one focusing on obtaining the financial performance variables and using them to perform a usual approach of ratio analysis. After the identification of capital structure variables, the next step is to perform relative analysis between variables as shown in table 1, which are derived from the concerned reviewed theories in literature. The final step is to suggest the process of model interpretation of the proposed theoretical framework. Thus, the process will result in tracing the practical implementation of generalized aspects of capital structure theories.

To elaborate the process in step two, it verifies following things. Firstly if tax information provided for a firm, the WACC (weighted average of cost of capital) decreases with an increase in debt financing and increases the value of the firm; secondly debt mix will not affect the Market value of the firm and the cost of capital and  in the other way market value of the firm is affected by operating income alone; thirdly, for a firm at a right combination of equity and debt, market value of the firm is maximum, but beyond a limit the value of the firm will diminish; fourthly, market value of firm is determined by the present value of future earnings and ignores the capital structure patterns; fifthly stated that marginal benefit of additional amount of debt, off sets the cost of capital; sixthly stated that managers will follow the preferences regarding usage of capital funds as follows: internal funds, debt financing, and lastly through equity financing; seventhly there will be trade-off between leverage and profitability and also to be noted that the optimal capital structure results from compromising between various functions of options from conflicts of interests.

**Table 1.** Theoretical Framework

| THEORY | VARIABLES | DESCRIPTION |
|---|---|---|
| **NET INCOME THEORY**<br><br>If tax information provided, it states that WACC ($K_0$) decreases with an increase in Debt financing and increase in Value of the firm | 1. Market value of the firm<br>2. Debt component changes | 1. Debt component has an impact on the market value of the firm i.e., net income impacts the market value of the firm |
| **NET OPERATING INCOME THEORY**<br><br>Debt mix will not affect the Market value and cost of capital.<br><br>Market value of the firm is affected by operating income only. | 1. Market value of the firm<br>2. Operating income<br>3. Debt equity ratio | 1. EBIT has an impact on the market value of the firm.<br><br>2. Capital structure has no impact on market value of the firm. |
| **MM THEORY**<br><br>Capital structure is not a factor for Market value of the firm | 1. Market value of the firm<br>2. Estimation of future cash flows. | 1. In this, market value of the firm is determined by the Present value of future earnings. |
| **TRADE OFF THEORY**<br><br>Marginal benefit of additional amount of debt off sets the cost of capital<br><br>At right combination of equity and debt, Market value of the firm is Maximum.<br><br>Debt beyond a limit, the value of the firm decreases | 1. Rate of change in debt component<br>2. Rate of changes in cost.<br>3. Market value of the firm<br>4. Debt component | 1. Change in debt is inversely proportional to change in cost.<br><br>2. Optimal debt component is obtained at the peak market value of the firm. |

| **PECKING ORDER THEORY**<br><br>Mangers will follow preferences:<br><br>1. Retained earnings<br><br>2. Debt financing<br><br>3. Equity | 1. Changes in retained earning<br><br>2. Changes in debt and equity financing<br><br>3. Asymmetry of information analysis | 1. Rate of change in the patterns of retained earnings; debt; and equity. |
| --- | --- | --- |
| **AGENCY THEORY**<br><br>1. Tradeoff between leverage and profitability<br><br>2. Optimal capital structure results from compromising between various functions of options from conflict of interests | 1. Correlation between debt ratios and profitability ratios | 1. Increase in leverage leads to decrease in agency costs. |

### 3.1 Variables of Interest

### 3.1.1. Dependent Variables

Market value of the firm, provides investors and stakeholders with an indication of the company's overall value and is a reflection of investors' expectations of future earnings and growth.

Rate of change in expenses, measures the growth rate of a company's expenses over time and can help identify areas where cost-cutting measures may be necessary.

Return on assets (ROA), measures a company's profitability relative to its total assets, and can provide insight into a company's efficiency in generating profits from its assets.

Return on equity (ROE), measures a company's profitability relative to its shareholders' equity, and can provide insight into a company's ability to generate returns for its investors.

## 3.1.2 Independent Variables

Return on equity (ROE), measures a company's profitability relative to its shareholders' equity, and can provide insight into a company's ability to generate returns for its investors.

The use of the debt-to-equity ratio to analyse a company's financial health and to compare the leverage of different firms. A high debt-to-equity ratio implies that a firm is significantly reliant on debt to fund its operations, which may be a warning indication of financial danger. A low debt-to-equity ratio, on the other hand, indicates that a company's activities are predominantly funded by equity, which can be an indication of financial soundness.

Long-term debt ratio, measures a company's long-term debt relative to its total assets and can provide insight into the company's leverage and financial stability.

Earnings before interest and tax (EBIT), represents a company's operating profit before accounting for interest and taxes, and can provide insight into a company's profitability and efficiency in generating revenue.

Estimation of future revenues, involves forecasting future revenues based on historical data and market trends, and can provide insight into a company's growth potential.

Long-term debt to asset ratio, measures a company's long-term debt relative to its total assets, and can provide insight into the cq`        `ompany's leverage and financial stability.

Long-term debt to equity ratio, measures a company's long-term debt relative to its shareholders' equity, and can provide insight into the company's leverage and financial risk.

## 3.2 Hypotheses Development

The basic objective of present study is the identification and applicability of capital structure theories of the select banks. The following are the specific objectives set for the study:

1. To study and understand the capital structure theories.
2. To analyse the capital structure patterns of select banks; and
3. To suggest the capital structure variables in investment decisions.

For the study, based on the above theoretical framework and objectives of the study, following alternative hypotheses are drawn:

$H_{11}$: Rate of change in long term debt has a significant impact on Market value of firm;
$H_{12}$: Earnings Before Interest & Tax has a significant impact on Market value of firm;
$H_{13}$: Debt Equity Ratio has a significant impact on Market Value of the firm;
$H_{14}$: Estimation of future revenues has a significant impact on Market Value of the firm;
$H_{15}$: Rate of change in long term debt has a significant impact on Rate of change in expences;
$H_{16}$: Long term debt to assets ratio and Return on assets are positively correlated;
$H_{17}$: Long term debt to assets ratio and Return on equity are positively correlated; and
$H_{18}$: Long term debt to assets ratio and Long term debt to equity have significant impact on ROA.

## 3.3. Selection of Banks

The present study focuses on the capital structure of banks and its impact on investment decisions, with a specific emphasis on PSU bank index banks in India. The study's scope is limited to six banks, namely Bank of Baroda, Bank of Maharastra, Union Bank of India, Bank of India, Central Bank of India, and Punjab National Bank.

**Table 2.** Public sector Bank index (PSUBANK Index) in National stock Exchange (NSE)

| Symbol | Name of the PSU | Date of Incorporation |
|--------|-----------------|----------------------|
| **BANKBARODA** | Bank of Baroda | July 20, 1908 |
| **BANKINDIA** | Bank of India | Sep 07, 1906 |
| **CANBK** | Canara Bank | July 01,1906 |
| **CENTRALBK** | Central Bank of India | 1911 |
| **INDIANBK** | Indian Bank | March 05, 1907 |
| **IOB** | Indian Overseas Bank | February 10, 1937 |
| **MAHABANK** | Bank of Maharashtra | September 16, 1935 |
| **PNB** | Punjab National Bank | April 12, 1895 |
| **PSB** | Punjab and Sind Bank | June 24, 1908 |
| **SBIN** | State Bank of India | June 02, 1806 |
| **UCOBANK** | UCO Bank | January 06, 1943 |
| **UNIONBANK** | Union Bank of India | November 11, 1919 |

These banks were selected from a population of 12 PSU banks listed on the National Stock Exchange of India. The sample selection was carried out using a systematic random sampling technique, which ensures that each bank in the population had an equal chance of being selected for the study. By limiting the study to PSU banks, the research aims to provide insights into the unique capital structures of government-owned banks in India and how these structures impact investment decisions. The findings of the study may not be generalizable to private or foreign banks, as they may have different capital structures and investment patterns.

## 4. Theories in Practice

### 4.1. Empirical Model

Though, every theory considered in the framework emphasize for a distinct relationship between the variables of interest, identifying the linear relationship from the observed values of the variables is important to determine the impact and level of dependency. Hence statistical tools like linear regression are appropriate in evaluating the significance of the relationship.

$$MVF_t = \beta_{0t} + \beta_{1t}RTD + \beta_{2t}EBIT + \beta_{3t}DER + \beta_{4t}ETFR + \epsilon_t \qquad (1)$$

$$REX_t = \alpha_{0t} + \alpha_{1t}RTD \qquad (2)$$

$$ROA_t = \gamma_{0t} + \gamma_{1t}LTDA + \gamma_{2t}LTDE \qquad (3)$$

As, stated in the section 3, variables that impact market value of the firm (MVF), return on assets (ROA) and rate of change in expenses (REX) are assessed through the above stated empirical equations.

In equation 1, a pooled regression equation is taken as used in other similar problems [13], it assumes that the change in MVF is not due to a single factor. When there is a change in the level of debt, rate of change in debt (RTD) results in a positive value, implying to show an impact on the market value of the firm. Similarly, the investors tend to assess the rate of change in operating income (EBIT), debt to equity ratio (DER) and the possible future revenues (ETFR) before investment.

Equation 2 represents a simple linear regression equation, it assumes a positive value of rate of change in debt would result in decline in the rate of change in the expenses (REX) benefitting the firm.

It is also assumed that a Peak value of the market value of the firm (PMVF) is due to an optimum DER. To verify whether the firm follows the pecking order theory, comparing the rate of change in retained earnings (RRE), RTD and rate of change in equity (REQ) is performed. To finding the change in profitability ratio's ROA, return on equity (ROE) and rate of change in earnings per share (REPS), the impact variables, long term debt ratio (LTDR) and long-term debt-equity ratio (LTDE) are considered in the equation 3.

### 4.2. Data Analysis

For the purpose of analysis, secondary data sources were used. The necessary information to extract the variables of interest was obtained from annual reports of the selected banks, as well as other important financial data from different public domains such as the NSE website, RBI website, and respective select banks' official websites The period between 2011 and 2022 was chosen for a data analysis using statistical tools such as linear regression, t-test, ANOVA, F-test, and Spearman correlation coefficient. This investigation sought to determine how banks' capital structures and long-term performance related to one another. Because of the distinctive capital structure that banks have, which is greatly impacted by legislation and market situations, banks were selected for this study.

Descriptive statistics such as mean, standard error, maximum, and minimum provide valuable insights into the distribution and characteristics of a particular variable or dataset. The mean, for instance, provides an estimate of the central tendency of the data, while the standard error can be used to gauge the precision of the estimate. The maximum and minimum values, on the other hand, provide an indication of the range of values the variable can take, and can help identify outliers or extreme values. By calculating and reporting these descriptive statistics for select variables in a dataset, researchers can gain a better understanding of the data's distribution, variability, and characteristics.

Table 3.a and 3.b provide a comprehensive overview of the descriptive statistics for select banks, and present the data in a clear and organized manner. The tables highlight key measures such as mean, standard deviation, minimum, and maximum values for various financial variables, and provide a comparison of these measures across different banks.

The tables also provide a useful reference for further analysis, such as hypothesis testing or regression analysis, and can help guide investment decisions and strategies.

The capital structure of the bank and its performance over time were investigated using linear regression. Using this statistical method, we were able to see any patterns or trends in the data and gauge how strongly the two variables were related.The performance of banks with various capital structures is an example of a situation where the t-test was employed to compare the means of two separate groups. We were able to evaluate whether the performance difference was statistically significant thanks to this test.

The table 4 containing dependent and independent variables with R-squared values and p-values is an essential tool for understanding the relationship between variables and identifying significant predictors. The R-squared number indicates the percentage of the dependent variable's variation that can be explained by the independent variables, whereas the p-value indicates the likelihood that the observed findings might have happened by chance. The p-value is deemed statistically significant if it is more than the significance level, which is commonly set at 1%, 5%, or 10%.

To read the table, start by identifying the dependent variable and the independent variables being analyzed. The R-squared value provides a measure of how much of the variation in the dependent variable can be explained by the independent variables, with higher values indicating a stronger relationship. The p-value indicates the significance of the relationship between each independent variable and the dependent variable, with values below the significance level indicating a statistically significant relationship. It is important to note that a low p-value does not necessarily mean that the relationship is strong or meaningful, and that other factors such as the direction and magnitude of the relationship should also be considered.

Table 5 contains the peak value of market value of the firm for each bank during the period along with the number of times the pecking order followed.

The strength and direction of the association between two non-normally distributed variables were finally determined using the Spearman correlation coefficient. Using this approach, we were able to assess if the capital structure of the bank and its financial performance were significantly related.

**Table 3.a.** Descriptive Statistics of the select variables of Bank of Baroda, Punjab National Bank, Bank of India

| | BOB | | | | PNB | | | | BOI | | | |
|---|---|---|---|---|---|---|---|---|---|---|---|---|
| | Mean | Standard Error | Minimum | Maximum | Mean | Standard Error | Minimum | Maximum | Mean | Standard Error | Minimum | Maximum |
| MVF | 2.63E+08 | 5113014 | 53993949 | 6E+08 | 1.76E+08 | 1723728 | 4393441 | 4.05E+08 | 1.76E+08 | 1608953 | 7708250 | 2.52E+08 |
| RTD | 0.106807 | 0.07923 | -0.29144 | 0.7225 | 0.240291 | 0.358341 | -0.46397 | 5.729347 | -0.66229 | 0.358341 | -0.66229 | 4.128571 |
| EBIT | 46954103 | 13557627 | -5.1E+07 | 1.21E+08 | 9044908 | 1528630 | -1.8E+08 | 82434400 | 9844908 | 1528630 | -7.6E+07 | 90932263 |
| DER | 0.301094 | 0.012948 | 0.24839 | 0.40426 | 0.020117 | 0.003953 | 0.006212 | 0.765003 | 0.020117 | 0.003953 | 0.002033 | 0.046067 |
| ETFR | 32126945 | 8566997 | -591216 | 69745133 | 18457999 | 1113156 | -1.1E+08 | 79507100 | 9844908 | 1113156 | -6.1E+07 | 56753471 |
| REX | 0.15381 | 0.044511 | 0 | 0.611429 | 0.078274 | 0.026369 | -0.14318 | 0.684471 | 0.054486 | 0.026369 | -0.04775 | 0.250865 |
| RRE | 0.156827 | 0.079634 | -0.31691 | 0.768752 | 0.215276 | 0.193057 | -0.02458 | 0.446353 | 0.00789 | 0.193057 | -0.593 | 2.20605 |
| REQ | 0.098434 | 0.06635 | 0 | 0.746565 | 0.193252 | 0.167575 | -0.02143 | 0.448648 | 0.026279 | 0.167575 | -0.45115 | 1.384156 |
| LTDE | 0.232077 | 0.008088 | 0.199099 | 0.286471 | 0.019559 | 0.003786 | 0.006212 | 0.335689 | 0.019559 | 0.003786 | 0.000029 | 0.044038 |
| LTDA | 0.301094 | 0.012948 | 0.24839 | 0.40426 | 0.020117 | 0.003953 | 0.006212 | 0.505319 | 0.020117 | 0.003953 | 0.002033 | 0.046067 |
| LTIA | 0.034288 | 0.01562 | 0.0155 | 0.205672 | 0.001176 | 0.000173 | 0.006251 | 2.959977 | 0.002033 | 0.000173 | -0.00018 | 0.007217 |
| ROA | 0.01257 | 0.009416 | -0.00733 | 0.114818 | 7.51E-05 | 0.001958 | 0.00367 | 0.71629 | 0.00789 | 0.001958 | -0.01634 | 0.14448 |
| ROE | 0.064965 | 0.027076 | -0.12122 | 0.190644 | 0.023989 | 0.0282 | -0.01615 | 0.199675 | 0.023989 | 0.0282 | -0.01634 | 0.14448 |
| REPS | -0.22305 | 0.542893 | -2.28257 | 5.048447 | -0.45352 | 0.313205 | -15.2708 | 4.58141 | -0.45352 | 0.313205 | -3.11688 | 0.652745 |

**Table 3.b.** Descriptive Statistics of the select variables of Union bank of India, Bank of Maharashtra, Central Bank of India.

| | UBI | | | | BOM | | | | CBI | | | |
|---|---|---|---|---|---|---|---|---|---|---|---|---|
| | Mean | Standard Error | Minimum | Maximum | Mean | Standard Error | Minimum | Maximum | Mean | Standard Error | Minimum | Maximum |
| MVF | 1.89E+08 | 18091946 | 72674282 | 2.73E+08 | 2.73E+08 | 49175395 | 18812150 | 1.37E+08 | 1.06E+08 | 13329606 | 50745276 | 1.93E+08 |
| RTD | 0.10149 | 0.115959 | -0.44873 | 0.953408 | 0.092135 | 0.109999 | -0.50515 | 0.898347 | -0.04046 | 0.03374 | -0.26162 | 0.151417 |
| EBIT | 14589789 | 11886327 | -5.9E+07 | 1.01E+08 | 8465753 | 6825030 | -4.5E+07 | 61846932 | -1.4E+07 | 10452009 | -8.1E+07 | 21116200 |
| DER | 0.417098 | 0.032529 | 0.264673 | 0.623973 | 0.449957 | 0.046279 | 0.258483 | 0.699566 | 0.312233 | 0.034946 | 0.113944 | 0.482162 |
| ETFR | 5901011 | 6989972 | -4.1E+07 | 36200100 | 4988202 | 3592826 | -0.05309 | 28806772 | -1.6E+07 | 8130030 | -6.5E+07 | 17418867 |
| REX | 0.182136 | 0.113992 | -0.00248 | 1.412289 | 0.077719 | 0.02993 | -0.58092 | 0.333647 | 0.054486 | 0.02596 | -0.06223 | 0.224081 |
| RRE | 0.16765 | 0.073473 | -0.13559 | 0.911425 | 0.129548 | 0.125548 | -0.41561 | 0.385431 | 0.00789 | 0.066401 | -0.60285 | 0.346061 |
| REQ | 0.172695 | 0.073218 | -0.13154 | 0.904646 | 0.148884 | 0.088012 | 0.205392 | 0.841615 | 0.026279 | 0.090976 | -0.54744 | 0.306111 |
| LTDE | 0.290306 | 0.016001 | 0.209282 | 0.384226 | 0.302599 | 0.028288 | 0.258483 | 0.092566 | 0.231867 | 0.0208 | 0.102289 | 0.32531 |
| LTDA | 0.341039 | 0.032529 | 0.264673 | 0.623973 | 0.449957 | 0.002909 | 0.258483 | 0.034014 | 0.312233 | 0.034946 | 0.113944 | 0.482162 |
| LTIA | 0.212057 | 0.001096 | 0.015711 | 0.027409 | 0.022438 | 0.022438 | 0.00265 | 0.034014 | 0.019135 | 0.019135 | 0.008102 | 0.006695 |
| ROA | 0.001762 | 0.001724 | -0.01061 | 0.008801 | 0.00265 | 0.002909 | 0.00265 | 0.034014 | -0.04867 | 0.03878 | -0.47108 | 0.006038 |
| ROE | 0.028752 | 0.033769 | -0.23999 | 0.169987 | -0.00824 | 0.085258 | -0.80965 | 0.464034 | -0.83651 | 0.660458 | -8.03281 | 0.062286 |
| REPS | -1.00976 | 0.765647 | -9.28091 | 0.738255 | -1.01096 | 1.008921 | -11.9528 | 0.901099 | -0.63838 | 0.334899 | -2.79531 | 0.970282 |

Table 4. Analysis for Linear relationship between the variables of interest of the sample

**DV: Market value of the firm**

| | | BOB | PNB | BOI | UBI | BOM | CBI |
|---|---|---|---|---|---|---|---|
| RTD | R square | 0.095832875 | 0.114931148 | 0.132593225 | 0.280466884 | 0.015584833 | 0.494918 |
| | P-VALUE | 0.327495 | 0.280037 | 0.244572 | 0.076594* | 0.699074 | 0.010685*** |
| EBIT | R square | 0.001518 | 0.026339 | 0.129203 | 0.403048 | 0.001 | 0.119719 |
| | P-VALUE | 0.904299 | 0.614307 | 0.251151 | 0.026564** | 0.922301 | 0.270595 |
| DER | R square | 0.062063 | 0.164409 | 0.049008 | 0.411236 | 0.298004 | 0.055369 |
| | P-VALUE | 0.43490105 | 0.19998 | 0.489268 | 0.024613** | 0.066347* | 0.4616 |
| ETFR | R square | 0.047537 | 0.045821 | 0.289124 | 0.000816 | 0.022813 | 0.020618 |
| | P-VALUE | 0.496029 | 0.5041 | 0.07137* | 0.929774 | 0.639381 | 0.656168 |

**DV: Rate of Change in Expenses**

| | | BOB | PNB | BOI | UBI | BOM | CBI |
|---|---|---|---|---|---|---|---|
| RTD | R square | 0.520447 | 0.003793 | 0.01728 | 0.41378 | 0.013698 | 0.195847 |
| | P-VALUE | 0.00809*** | 0.849196 | 0.683838 | 0.024032** | 0.717175 | 0.149682 |

**DV: Return on Assets**

| | | BOB | PNB | BOI | UBI | BOM | CBI |
|---|---|---|---|---|---|---|---|
| LTDA | R square | 0.9768 | 0.8950 | 0.4479 | 0.6280 | 0.2925 | 0.2519 |
| | p- Value | 1.6416E-09*** | 3.28738E-06*** | 0.01730** | 0.002112*** | 0.06943* | 0.09640* |

**DV: Return on Equity**

| | | BOB | PNB | BOI | UBI | BOM | CBI |
|---|---|---|---|---|---|---|---|
| LTDA | R square | 0.320335 | 0.13889 | 0.515578 | 0.054075 | 0.00892 | 0.170283 |
| | p- Value | 0.055079* | 0.232836 | 0.00854*** | 0.467035 | 0.770322 | 0.182486 |

Robust standard errors in parentheses, *** p<0.01, ** p<0.05, * p<0.1.

Table 5. Peak value of MVF and pecking order verification

| Banks | IV: Debt-Equity Ratio | | | Pecking Order |
|---|---|---|---|---|
| | Lag Effect | Contemporaneous Effect | Peak Value | No of Years Followed (For n=12) |
| BOB | 0.35873644 | 0.303125567 | 599877992 | 0 |
| PNB | 0.16233064 | 0.194078123 | 404735863 | 0 |
| BOI | --- | 0.046066591 | 251572068 | 0 |
| UBI | 0.27271712 | 0.264672626 | 272706405 | 0 |
| BOM | 0.25848276 | 0.292948899 | 136953840 | 2 |
| CBI | 0.38147576 | 0.32588204 | 193071466 | 0 |

Table 6.a to 6.g present the Spearman correlation coefficient of variables of interest of select banks. The tables indicate the strength and direction of the relationship between the variables, which is important in understanding the interplay of these variables and how they affect the overall performance of the banks. By examining the correlation coefficients, we can identify which variables have a positive or negative relationship and to what extent.

Overall, this data analysis provided valuable insights into the relationship between the capital structure of banks and their performance over time. The statistical tools used allowed us to identify any trends, patterns, and significant differences in performance across different groups. Understanding the relationship between a bank's capital structure and its performance is critical for investors and financial institutions, as it can help them make informed investment decisions and manage risk effectively.

**Table 6a Correlation matrix for the variables of Central Bank of India**

| | MVF | REX | RDT | EBIT | DER | ETFR | RRE | REQ | LTDR | LTDE | LDTA | ROA | ROE | REPS |
|---|---|---|---|---|---|---|---|---|---|---|---|---|---|---|
| MVF | 1 | | | | | | | | | | | | | |
| REX | -0.23218 | 1 | | | | | | | | | | | | |
| RDT | -0.7035 | 0.442545 | 1 | | | | | | | | | | | |
| EBIT | -0.346 | 0.31681 | 0.319744 | 1 | | | | | | | | | | |
| DER | -0.23531 | 0.293682 | 0.692095 | 0.113071 | 1 | | | | | | | | | |
| ETFR | -0.14359 | 0.376614 | 0.410984 | 0.768158 | 0.484749 | 1 | | | | | | | | |
| RRE | 0.195823 | 0.461934 | -0.19167 | 0.123134 | -0.31835 | 0.125896 | 1 | | | | | | | |
| REQ | 0.17836 | 0.445582 | -0.23077 | 0.023812 | -0.45434 | -0.02624 | 0.968358 | 1 | | | | | | |
| LTDR | -0.22902 | 0.282121 | 0.67755 | 0.062733 | 0.996707 | 0.44342 | -0.3007 | -0.43793 | 1 | | | | | |
| LTDE | -0.23531 | 0.293682 | 0.692095 | 0.113071 | 0.987601 | 0.484749 | -0.31835 | -0.45434 | 0.996707 | 1 | | | | |
| LDTA | -0.50861 | 0.22267 | 0.783184 | 0.355047 | 0.778155 | 0.621733 | -0.25109 | -0.39785 | 0.777945 | 0.778155 | 1 | | | |
| ROA | -0.02516 | 0.271071 | 0.648861 | 0.099284 | 0.410486 | 0.341485 | 0.014519 | 0.010111 | 0.415956 | 0.410486 | 0.501871 | 1 | | |
| ROE | -0.02455 | 0.27432 | 0.609728 | 0.093329 | 0.412654 | 0.340619 | 0.014656 | 0.010487 | 0.415998 | 0.412654 | 0.499794 | 0.999967 | 1 | |
| REPS | 0.027348 | -0.5297 | -0.14765 | -0.30557 | -0.17712 | -0.30024 | 0.022081 | 0.032553 | -0.1382 | -0.17712 | -0.0867 | -0.14777 | -0.151 | 1 |

**Table 6.b Correlation matrix for the variables of Bank of Maharashtra.**

| | MVF | REX | RDT | EBIT | DER | ETFR | RRE | REQ | LTDR | LTDE | LDTA | ROA | ROE | REPS |
|---|---|---|---|---|---|---|---|---|---|---|---|---|---|---|
| MVF | 1 | | | | | | | | | | | | | |
| REX | -0.30726 | 1 | | | | | | | | | | | | |
| RDT | -0.12484 | -0.11704 | 1 | | | | | | | | | | | |
| EBIT | 0.031615 | 0.10739 | 0.429699 | 1 | | | | | | | | | | |
| DER | -0.5459 | -0.27258 | -0.18142 | 0.044542 | 1 | | | | | | | | | |
| ETFR | -0.15104 | 0.210927 | 0.051361 | 0.534466 | 0.201843 | 1 | | | | | | | | |
| RRE | 0.091667 | 0.25148 | 0.69993 | 0.550104 | -0.05371 | 0.277825 | 1 | | | | | | | |
| REQ | 0.02464 | -0.32471 | 0.665361 | 0.370641 | 0.026024 | -0.03461 | 0.932067 | 1 | | | | | | |
| LTDR | 0.415729 | -0.12591 | 0.12228 | -0.00974 | -0.28811 | 0.17959 | -0.11312 | -0.25509 | 1 | | | | | |
| LTDE | 0.388572 | -0.1337 | 0.126015 | -0.02033 | -0.25766 | 0.177532 | -0.1911 | -0.27747 | 0.997557 | 1 | | | | |
| LDTA | 0.250775 | -0.22806 | 0.141261 | 0.11404 | 0.052151 | 0.131182 | -0.12788 | -0.16799 | 0.688667 | 0.704424 | 1 | | | |
| ROA | 0.250775 | -0.22806 | 0.141261 | 0.11404 | 0.052151 | 0.131182 | -0.12788 | -0.16799 | 0.688667 | 0.704424 | 0.88743 | 1 | | |
| ROE | 0.195664 | -0.02891 | 0.508251 | 0.955425 | -0.05496 | 0.535264 | 0.70871 | 0.523677 | 0.112925 | 0.094444 | 0.173933 | 0.173933 | 1 | |
| REPS | 0.129743 | -0.10297 | -0.01097 | 0.152457 | -0.26141 | 0.002067 | 0.264047 | 0.227213 | -0.47561 | -0.51346 | -0.39268 | -0.39268 | 0.159098 | 1 |

**Table 6.c** Correlation matrix for the variables of Bank of India

| | MVF | REX | RDT | EBIT | DER | ETFR | RRE | REQ | LTDR | LTDE | LDTA | ROA | ROE | REPS |
|---|---|---|---|---|---|---|---|---|---|---|---|---|---|---|
| MVF | 1 | | | | | | | | | | | | | |
| REX | -0.31026 | 1 | | | | | | | | | | | | |
| RDT | 0.364134 | -0.13145 | 1 | | | | | | | | | | | |
| EBIT | 0.359448 | -0.84591 | 0.10431 | 1 | | | | | | | | | | |
| DER | 0.221378 | -0.0311 | -0.00052 | -0.22879 | 1 | | | | | | | | | |
| ETFR | 0.537703 | -0.46664 | 0.176091 | 0.598687 | -0.56022 | 1 | | | | | | | | |
| RRE | -0.16679 | -0.15189 | -0.07645 | 0.001975 | -0.22069 | 0.12034 | 1 | | | | | | | |
| REQ | -0.31995 | -0.10154 | -0.2723 | -0.00205 | -0.21095 | 0.01287 | 0.970563 | 1 | | | | | | |
| LTDR | 0.215366 | -0.02593 | 0.003389 | 0.399909 | -0.56538 | -0.22125 | -0.22069 | -0.21155 | 1 | | | | | |
| LTDE | 0.221378 | -0.0311 | -0.00052 | -0.22879 | 0.998701 | -0.56022 | -0.22069 | -0.21095 | 0.999909 | 1 | | | | |
| LDTA | 0.237183 | -0.16992 | 0.004363 | -0.10136 | 0.95428 | -0.45617 | 0.010035 | 0.01465 | 0.954277 | 0.95428 | 1 | | | |
| ROA | 0.398187 | -0.00080 | 0.228917 | -0.21526 | 0.736933 | -0.14731 | -0.29352 | -0.38251 | 0.73843 | 0.736933 | 0.669243 | 1 | | |
| ROE | 0.487719 | -0.02693 | 0.15056 | -0.22728 | 0.718637 | -0.07771 | -0.12647 | -0.22364 | 0.717354 | 0.718637 | 0.671402 | 0.960762 | 1 | |
| REPS | 0.414205 | -0.03354 | -0.4826 | 0.044204 | -0.02388 | 0.399846 | 0.113846 | 0.088091 | -0.02983 | -0.02388 | -0.02662 | 0.240005 | 0.425486 | 1 |

**Table 6.d** Correlation matrix for the variables of Union Bank of India

| | MVF | REX | RDT | EBIT | DER | ETFR | RRE | REQ | LTDR | LTDE | LDTA | ROA | ROE | REPS |
|---|---|---|---|---|---|---|---|---|---|---|---|---|---|---|
| MVF | 1 | | | | | | | | | | | | | |
| REX | 0.465766 | 1 | | | | | | | | | | | | |
| RDT | 0.529591 | 0.643257 | 1 | | | | | | | | | | | |
| EBIT | 0.634861 | 0.181203 | 0.099642 | 1 | | | | | | | | | | |
| DER | -0.64128 | -0.39365 | -0.28354 | -0.47646 | 1 | | | | | | | | | |
| ETFR | 0.028567 | -0.12969 | -0.21022 | 0.707892 | 0.079565 | 1 | | | | | | | | |
| RRE | 0.516562 | 0.913254 | 0.648352 | 0.116224 | -0.63025 | -0.3191 | 1 | | | | | | | |
| REQ | 0.503557 | 0.900437 | 0.61278 | 0.08886 | -0.63065 | -0.37159 | 0.997916 | 1 | | | | | | |
| LTDR | -0.67472 | -0.64165 | -0.30209 | -0.48342 | 0.996448 | 0.08521 | -0.64133 | -0.62295 | 1 | | | | | |
| LTDE | -0.64172 | -0.39365 | -0.28354 | -0.47646 | 0.951068 | 0.079565 | -0.63025 | -0.65025 | 0.969375 | 1 | | | | |
| LDTA | 0.250492 | 0.07508 | -0.0951 | 0.784751 | -0.20365 | 0.64887 | -0.0344 | -0.65942 | -0.17373 | -0.20365 | 1 | | | |
| ROA | 0.250092 | 0.072685 | -0.07465 | 0.775236 | -0.23254 | 0.615456 | -0.09135 | -0.86677 | -0.20213 | -0.23254 | -0.03794 | 1 | | |
| ROE | 0.241762 | -0.00943 | | | | | | -0.08102 | -0.45597 | | -0.06645 | 0.994449 | 1 | |
| REPS | 0.283719 | -0.01943 | 0.101006 | 0.635914 | -0.4821 | 0.307035 | 0.011436 | -0.01164 | | | -0.32687 | 0.705498 | 0.749222 | 1 |

**Table 6.e** Correlation matrix for the variables of Bank of Baroda

| | MVF | REX | RTD | EBIT | DER | ETFR | RRE | REQ | LTDR | LTDE | LDTA | ROA | ROE | REPS |
|---|---|---|---|---|---|---|---|---|---|---|---|---|---|---|
| MVF | 1 | | | | | | | | | | | | | |
| REX | -0.27326 | 1 | | | | | | | | | | | | |
| RTD | -0.30957 | 0.72142 | 1 | | | | | | | | | | | |
| EBIT | 0.038967 | -0.32294 | -0.13864 | 1 | | | | | | | | | | |
| DER | -0.24912 | -0.50669 | -0.04392 | 0.546083 | 1 | | | | | | | | | |
| ETFR | 0.21803 | -0.07093 | -0.17901 | 0.485188 | -0.09704 | 1 | | | | | | | | |
| RRE | 0.136029 | 0.373643 | -0.2211 | 0.228408 | -0.33227 | 0.166181 | 1 | | | | | | | |
| REQ | -0.18978 | 0.92054 | 0.758722 | -0.34101 | -0.36332 | -0.32512 | 0.320177 | 1 | | | | | | |
| LTDR | -0.26922 | -0.26529 | 0.250083 | 0.517837 | 0.944931 | -0.11409 | -0.34505 | -0.11 | 1 | | | | | |
| LTDE | -0.24912 | -0.50669 | -0.04392 | 0.546083 | 1 | -0.09704 | -0.33227 | -0.36332 | 0.944931 | 1 | | | | |
| LDTA | -0.35995 | -0.07519 | -0.1016 | 0.201806 | 0.26173 | -0.35036 | 0.151583 | -0.07408 | 0.245946 | 0.26173 | 1 | | | |
| ROA | -0.41147 | -0.09635 | -0.14316 | 0.292987 | 0.287442 | -0.28348 | 0.202129 | -0.11351 | 0.249854 | 0.287442 | 0.988372 | 1 | | |
| ROE | -0.50421 | -0.23136 | -0.17335 | 0.77457 | 0.565981 | 0.194725 | 0.226623 | -0.27591 | 0.466444 | 0.565981 | 0.417573 | 0.540857 | 1 | |
| REPS | 0.321808 | -0.10962 | -0.22393 | 0.797007 | 0.310455 | 0.485519 | 0.603723 | -0.10673 | 0.311956 | 0.310455 | 0.056178 | 0.111431 | 0.454023 | 1 |

**Table 6.f** Correlation matrix for the variables of Punjab National Bank

| | MVF | REX | RDT | EBIT | DER | ETFR | RRE | REQ | LTDR | LTDE | LDTA | ROA | ROE | REPS |
|---|---|---|---|---|---|---|---|---|---|---|---|---|---|---|
| MVF | 1 | | | | | | | | | | | | | |
| REX | 0.236118 | 1 | | | | | | | | | | | | |
| RDT | 0.339015 | 0.061589 | 1 | | | | | | | | | | | |
| EBIT | 0.162292 | 0.086203 | -0.63417 | 1 | | | | | | | | | | |
| DER | -0.40547 | 0.09306 | 0.068778 | -0.15789 | 1 | | | | | | | | | |
| ETFR | 0.214059 | 0.334068 | -0.48368 | 0.654637 | 0.216635 | 1 | | | | | | | | |
| RRE | 0.105563 | 0.478137 | 0.01651 | 0.18941 | 0.035493 | 0.424088 | 1 | | | | | | | |
| REQ | 0.114992 | 0.478517 | 0.014189 | 0.195379 | 0.029572 | 0.430285 | 0.999858 | 1 | | | | | | |
| LTDR | -0.17091 | 0.071113 | 0.358169 | -0.35398 | 0.528995 | 0.027468 | -0.09518 | -0.09902 | 1 | | | | | |
| LTDE | -0.22537 | 0.031099 | 0.39103 | -0.39569 | 0.500502 | -0.08163 | -0.15992 | -0.16476 | 0.990586 | 1 | | | | |
| LDTA | -0.55531 | 0.069994 | -0.17927 | -0.54271 | 0.446637 | -0.00757 | 0.003174 | -0.00244 | 0.186121 | 0.173292 | 1 | | | |
| ROA | -0.58312 | 0.076769 | -0.18522 | -0.42685 | 0.666279 | 0.076768 | 0.045718 | 0.039245 | 0.165051 | 0.145973 | 0.946047 | 1 | | |
| ROE | -0.38729 | -0.08273 | 0.153799 | -0.21908 | 0.398477 | -0.03124 | 0.280644 | 0.267882 | 0.344123 | 0.37268 | 0.285207 | 0.347864 | 1 | |
| REPS | 0.119009 | -0.12639 | 0.164688 | -0.06546 | 0.031286 | 0.099642 | 0.462947 | 0.460811 | -0.06621 | -0.0761 | 0.095608 | 0.121799 | 0.649922 | 1 |

### 4.3. Interpretation

Bank of Baroda followed the traditional theory of capital structure, which suggests that firms should aim to have a balanced mix of debt and equity to maximize their value. They also followed the agency cost theory, which highlights the conflicts of interest between managers and shareholders and suggests that debt can be used to align these interests. The bank did not follow the net income theory, NOI theory, MM theory, or pecking order theory.

Punjab National Bank followed the agency costs theory, which is consistent with the idea that debt can be used to align the interests of managers and shareholders. Bank of India nearly followed MM theory, which suggests that the value of a firm is independent of its capital structure. The bank also thoroughly followed agency cost theory.

Union Bank of India followed net income theory and more significantly, the net operating income theory, which suggests that a firm should use retained earnings to finance its investment opportunities. They also followed the traditional theory and agency costs theory but did not follow the pecking order theory in practice.

Bank of Maharashtra nearly followed the net income theory and pecking order theory for two times in the given period. The net income theory suggests that a firm should use retained earnings to finance its investment opportunities, while the pecking order theory suggests that firms should prioritize internal funds and debt before issuing equity. Central Bank of India followed the net income theory and partially followed agency costs theory.

Overall, the different capital structure theories followed by these banks imply that each bank has a unique approach to financing its investments and managing its capital structure. By analyzing the theories followed by these banks, we can gain insight into their decision-making processes and how they balance the costs and benefits of different financing options.

Bank of Maharashtra nearly followed the net income theory and pecking order theory for two times in the given period. The net income theory suggests that a firm should use retained earnings to finance its investment opportunities, while the pecking order theory suggests that firms should prioritize internal funds and debt before issuing equity. Central Bank of India followed the net income theory and partially followed agency costs theory.

Overall, the different capital structure theories followed by these banks imply that each bank has a unique approach to financing its investments and managing its capital structure. By analyzing the theories followed by these banks, we can gain insight into their decision-making processes and how they balance the costs and benefits of different financing options.

### 4.4. Recommendation to investors

Following benchmarks are derived from prior research for the information of investors, which aids the investors in comparing the actuals with the standards and incorporating in their analysis process to place their funds as an effective investment:

**Table 7: Standard Benchmark Values for different analysis**

| Analysis | Standard Benchmark values |
|---|---|
| **Fundamental analysis** | |
| Revenue Growth Rate | Above 10% annually is considered good [23] |
| Net Profit Margin | Above 15% is considered good [24] |
| Return on equity | Above 15% is considered good [25] |
| Debt to equity | Below 1 is considered good [26] |
| **Technical analysis** | |
| Moving average | 50 day and 200 day moving averages [27] |
| Relative strength index | Overbought market- above 70 |
| | Oversold market – below 30 [28] |

**Quantitative analysis**

| | |
|---|---|
| Price to Earnings Ratio | Below 20 is considered good [29] |
| Price to sales Ratio | Below 2 is considered good [30] |
| Dividend Yield | Above 2% is considered good [31] |
| Price to earnings growth | Below 1 is considered good [32] |

**Environmental, Social and Governance analysis**

| | |
|---|---|
| Carbon foot print | Low carbon foot print is better [33] |
| Labour practices | Providing fair wages benefits and safe working conditions [33] |
| Board diversity | Diverse board of directors [33] |

As previously said, analysis aids investors in several ways to make profitable investments. In a similar vein, investors who follow capital structure variable benchmarks are better equipped to comprehend the nature and rate of change of such capital structure variables.

Following are some of the benchmarks of capital structure variables:

**Table 8: Standard Benchmark Values for Capital structure variables**

| Capital structure variables | Standard Benchmark values |
|---|---|
| Debt to Equity Ratio | Below 1.5 is considered good [34] |
| Debt to Asset Ratio | Below 0.6 is considered good [35] |
| Interest Coverage Ratio | Above 1.5 is considered good [36] |
| Debt service coverage Ratio | Above 1 is considered good [37] |
| Weighted average cost of capital | Below ROI is considered good [38] |

It is important to note that these benchmarks may vary depending on the industry, company size, and other factors. Retail investors should also consider analyzing these ratios in combination with other factors and benchmarks to make informed investment decisions.

## 5. Conclusion

The study of capital structure theories in practice in banks is crucial because it provides insight into the financing decisions made by these institutions. The capital structure of a bank refers to the mix of debt and equity used to finance its operations, and the decisions made about this mix can have a significant impact on the bank's financial performance and risk profile. By understanding the different theories of capital structure, banks can make informed decisions about the optimal mix of capital to use to achieve their strategic objectives.

Capital structure behavior analysis, along with other financial performance analysis of firms, is essential for investors to make accurate investment decisions. Investors can use financial performance indicators, such as leverage ratios, interest coverage, and debt maturity structure, to identify potential risks and opportunities associated with a particular company. By analyzing a firm's capital structure behavior, investors can identify how the company is using its debt and equity and make informed decisions about investing in a particular company.

Investors can also benefit from understanding the different theories of capital structure, such as the trade-off theory, the pecking order theory, and the agency theory. These theories provide insight into the factors that influence a firm's capital structure decisions and can help investors identify potential risks associated with a company's financing decisions. For example, if a company has a high level of debt and a low level of equity, investors may be concerned about the company's ability to repay its debt and may avoid investing in that company.

In conclusion, the study of capital structure theories in practice in banks and other firms is critical for informed decision-making. Understanding the impact of debt and equity on a firm's financial performance and risk profile can aid banks in determining the optimal mix of capital to use to achieve their strategic objectives. Similarly, investors can benefit from analyzing a firm's capital structure behavior along with other financial performance indicators to make accurate investment decisions. By assessing a company's capital structure decisions and considering the different theories of capital structure, investors can identify potential risks and opportunities and make informed decisions about investing in a particular company.